# Statistical Compatibility, Refutational Information, and Acceptability

Alessandro Rovetta[1]

1. International Committee Against the Misuse of Statistical Significance (ICAMSS), Bovezzo, Italy

**Corresponding author**

Alessandro Rovetta
International Committee Against the Misuse of Statistical Significance
Bovezzo (BS), 25073, Italy
alessandrorovetta@icamss.com
+39 3927112808
https://orcid.org/0000-0002-4634-279X

**Summary**

This paper develops an interpretive framework for divergence P-values and S-values within a descriptive frequentist perspective. Statistical analysis is framed as operating within idealized worlds defined by a set of assumptions and a target hypothesis, where probabilities describe the behavior of data under the model but do not assign truth values to hypotheses. Within this view, P-values are interpreted as graded indices of compatibility between the observed result and the predictions generated by the assumed model; accordingly, small P-values should not be read as indicating logical impossibility or strict inconsistency of the model itself. Building on this distinction, the paper argues that practical inference requires moving beyond the internal logic of the model toward judgments of overall acceptability, which depend not only on data-model compatibility but also on multiple contextual considerations such as subject-matter knowledge, plausibility of assumptions, data quality, usefulness, and loss - all interpreted through the competence, intentions, perceptions, and moral values of the specific analyst. S-values are therefore interpreted not as evidence against the epistemic status of the model, but as a specific form of refutational information that contributes to the broader body of information used by the analyst to judge whether a model remains acceptable for an intended practical purpose. The paper also examines the linguistic and conceptual risks associated with the language of incompatibility, distinguishes probability from rarity, and clarifies different notions of surprise - including a possible definition of Shannon-type surprise, to be distinguished from Bayesian belief revision. Overall, the article proposes a more cautious and explicit interpretation of frequentist measures, centered on model-based description, analyst responsibility, and decision acceptability. These considerations mainly pertain to settings with accumulated, multi-source evidence; in contrast, individual studies lacking such evidentiary breadth should primarily aim to provide clear and transparent descriptions of the data-model relationship rather than attempt to resolve questions of practical acceptability. In this regard, the paper also discusses how Bayesian posterior intervals can be interpreted as ranges of parameter values reasonably compatible with the data and prior information, under the procedure specified by the analyst.

## INTRODUCTION

Recent attempts to improve statistical interpretation have increasingly emphasized continuous measures of compatibility, information, and surprise in place of dichotomous notions such as "statistical significance" and "non-significance" (Amrhein, Greenland, et al., 2019; Greenland et al., 2016; Mansournia et al., 2022; McShane et al., 2019; Rovetta, Piretta, et al., 2025; Wasserstein & Lazar, 2016). These developments represent an important improvement in statistical communication (Ting & Greenland, 2024), but they also raise further interpretive questions and challenges (Rovetta & Mansournia, 2025). In particular, all these terms may refer to conceptually different levels of statistical reasoning unless their meanings and domains of application are made explicit.

A major distinction developed in this paper is between quantities defined within an assumed statistical model and judgments made about that model in the broader real-world context. A P-value or S-value is computed within an idealized probabilistic system defined by a set of assumptions about the data-generating process and a target hypothesis about the parameter of interest. It therefore characterizes the relationship between the observed result and the predictions generated within that system. By contrast, deciding whether the model remains acceptable for a particular scientific or practical purpose requires leaving that conditional system and integrating the statistical result with external considerations such as subject-matter knowledge, plausibility of assumptions, data quality, causal information, usefulness, and the consequences of possible errors.

On this basis, the paper develops an interpretive framework centered on three related but distinct concepts: statistical compatibility, refutational information, and acceptability. Statistical compatibility describes the degree of agreement between an observed result and the predictions generated under a specific model with a regular reference distribution. Refutational information characterizes the extent to which the observed result departs from those predictions. Acceptability, by contrast, is a broader analyst-dependent judgment concerning whether the model remains adequate for an intended purpose after compatibility and refutational information have been considered together with relevant external evidence and decision consequences. The aim is not to introduce a new statistical procedure, but to clarify the interpretation and appropriate scope of existing frequentist quantities and their relation to broader inferential and decision processes.

The paper refers exclusively to divergence P-values as defined in the descriptive frequentist framework and distinguished from decision P-values by Greenland (Greenland, 2023b). Following Greenland's notation, A denotes the set of assumptions defining the working model, while $H_i$ denotes a specific target hypothesis under evaluation; together, A + $H_i$ defines the conditional probabilistic system under consideration. The discussion is restricted to cases in which the reference distribution of the test statistic satisfies the regularity conditions required for the compatibility interpretation developed below. Finally, the formulations proposed here are not intended to provide definitive truth evaluations of models or hypotheses, but to offer interpretive distinctions that may be useful in statistical practice (Box, 1976; Msaouel, 2023).

## 2. FREQUENTIST FRAMEWORKS: BASIC CONCEPTS

In frequentist frameworks, statistical analysis is grounded in a set of assumptions A that jointly specify the expected behavior of the phenomenon under study (Cox & Hinkley, 1979; Greenland, 2023b). These assumptions ideally encode all the relevant properties of the data-generating process (DGP), and they are combined with a target hypothesis $H_i$ to define the model A + $H_i$. In particular, A includes background conditions such as distributional forms of data and errors, degree of dependence, measurement properties, and all structural and causal features that define how data are mechanically produced; the target hypothesis $H_i$ specifies a particular "claim" about the parameter of interest (e.g., the average efficacy of a medical treatment in the defined population is zero). Different target hypotheses correspond to different model specifications; in some cases these define families of admissible models (e.g., composite hypotheses such as 'the average effect is greater than zero'), whereas in others they define fully specified models. For instance, one may consider the null model A + $H_0$ - where $H_0$ encodes the null hypothesis of zero effect - or the alternative non-null models A + $H_{i\neq0}$ (e.g., A + $H_1$ or A + $H_2$) - where $H_{i\neq0}$ encodes the target hypothesis of a specific (i) non-zero effect; each of these defines a *distinct*, *internally coherent* probabilistic system.

Crucially, these systems should *not* be interpreted as competing claims endowed with probabilities of being true, as the “traditional” frequentist frameworks do *not* assign probabilities to hypotheses or models. Instead, each model represents a conditional scenario: if the assumptions A and the specific hypothesis $H_i$ were true, then certain probabilistic statements about the data would follow. For this reason, it is useful to conceptualize these models as “small fictional worlds” defined by a *predefined* combination of assumptions and a specific target hypothesis; *within that world*, one can derive the distribution of possible data and the behavior of statistical procedures (Greenland, 2025). The analyst can thus “enter” each of these worlds and examine what would be expected to occur under those conditions - in the language of Cox and Hinkley (Cox & Hinkley, 1979), one treats the chosen model $A + H_i$ as *provisionally* true.

## 3. ON P-VALUES, COMPATIBILITY, AND INCOMPATIBILITY

The P-value can be defined as the probability of obtaining, under *the* specific model $A + H_i$, a statistical outcome (e.g., a test statistic) at least as extreme as the one observed (Wasserstein & Lazar, 2016). The reference is therefore always to a fully idealized world, not to the hypothesis in isolation. In common instances, this definition naturally supports an interpretation of the P-value as an index of *compatibility* - ranging from 0 to 1 - between the observed result and the predictions generated within that model; here, compatibility should be understood as a graded measure of *agreement*, *consonance,* or *consistency* between what is expected under $A + H_i$ (in terms of the sampling distribution of the statistic) and what is actually observed (Mansournia et al., 2022; Rovetta, Piretta, et al., 2025; Greenland, 2023b; Wasserstein et al., 2019; McShane et al., 2024; Rovetta, 2025; Hennig, 2024). P-values “close” to 1 correspond to outcomes that are “typical” under the assumed model, and thus indicate “high” compatibility; as the P-value decreases, the observed result becomes increasingly “atypical” relative to the model’s predictions, so that the degree of compatibility correspondingly decreases.

However, it would be incorrect to interpret “small” P-values as showing logical incompatibility, impossibility, or inconsistency with $H_i$. The model $A + H_i$ defines an internally coherent probabilistic system within which all realizations in the support of the distribution are, by construction, perfectly possible. The observed data cannot contradict the model in a strict logical sense, because the model is provisionally assumed as the generating mechanism for the purpose of the calculation (Cox & Hinkley, 1979; Greenland et al., 2016).

Statements about “incompatibility” are conventions with their own merits, but involve non-negligible interpretative risks in light of the *inversion fallacies* that pervade frequentist statistics (Greenland et al., 2016; Rovetta & Mansournia, 2025). Etymologically, ‘incompatibility’ derives from the Latin ‘incompatibilis’, denoting the *inability* of two (or more) things to co-exist or be brought together - a meaning that naturally conveys *total* conflict rather than a matter of degree (Douglas Harper, n.d.). Such an original meaning has been preserved in ordinary language (“The Origin of Incompatibility: From Past to Present,” n.d.). Even in probability theory and in standard statistical usage, it is more common to reserve ‘incompatibility’ for genuine structural impossibility: situations in which a model forbids an outcome or pattern outright, so that the corresponding event lies outside the model's support (Philip B Stark, 2019). In the special case of discrete models this often coincides with the model assigning probability zero, and for mutually exclusive (or disjoint) events it means they cannot occur together. The common underlying idea is thus binary: either an event is possible under the model or it is not.

The temptation to speak of “degrees of incompatibility” often arises because we want a graded scale to replace the older dichotomies of “significant” versus “not significant”. But importing graded language into a term that is frequently used dichotomically risks conceptual muddle. Consider the parallel with “insignificance”. Historically, some authors including Neyman have used graded labels for statistical significance (for example, “highly significant”, “approximately significant”, or “marginally significant”) (Gelman, 2013; Neyman, 1977). Yet, at least at the level of ordinary phrasing, it is less natural to apply an analogous graded vocabulary to the “not significant” side - that is, to “insignificance” and related variants - unless one is prepared to shift the term’s meaning away from simple negation and toward a graded quantity. This is a linguistic expectation that can be checked empirically: in a PubMed title and abstract corpus analysis since 1926, reported in the Supplement, the conditional frequency of the examined graded modifiers was about 12.3-fold higher on the “significant/significance” side than on the “insignificance” or "not/non-significant" side. Such a result strongly agrees with the hypothesis that "insignificant" is naturally read as the negation or absence of "significant", not as a quantity with its own continuum. As an interesting aside, mentions of "significance" were 18 times as frequent as mentions of "non-significance," consistent with the well-known "significance filter" (Van Zwet & Cator, 2021).

This matters even more when we move across inferential cultures. In Bayesian contexts, one may speak about conflict between prior and likelihood, between model components, or about prior-data conflict and model checking (Evans & Moshonov, 2006). Speaking of incompatibility of hypotheses with data risks conflating distinct ideas: tail-area rarity under a frequentist reference model and posterior predictive misfit in Bayesian checking. These are not the same object, and a single word should not be made to carry them all unless we carefully define it. For these reasons, contrary to what I myself have done in other works (Rovetta & Mansournia, 2024), I argue it is more "appropriate" (less risky) to frame discussions in terms of graded compatibility rather than graded incompatibility. Within a given model, outcomes are compatible insofar as they are possible under the model, but they can be more or less in accord with what the model makes "typical". The P-value, in this view, can be described as one particular measure of extremeness (e.g., relative to a test statistic and a reference distribution) and thus as an indirect indicator of how "unusual" the computed result looks under the assumed model.

Moreover, we increasingly see new dichotomies emerging under the banner of "compatibility language". For example, some writings describe parameter values outside a so-called 'compatibility interval' as incompatible, and those inside as compatible (Rovetta & Mansournia, 2025). This reproduces almost verbatim the "inside-outside" dichotomy of confidence intervals, with 'compatible/incompatible' simply replacing 'accepted/rejected' or 'significant/not significant' (Amrhein et al., 2017). If we then also speak of degrees of incompatibility, we risk creating a conceptual hybrid that is simultaneously graded and dichotomous, depending on rhetorical needs. That is precisely the kind of language instability that invites misinterpretation and overconfident conclusions.

## 4. S-VALUES, REFUTATIONAL INFORMATION, AND ACCEPTABILITY

The practical use of frequentist results requires a step that lies outside those "fictional small worlds". The analyst should exit the provisional abstraction defined by A and $H_i$, so that the key question is no longer how the data behave within a hypothetical world, but to what extent that hypothetical world should be considered *acceptable* in light of the broader evidential and decision context (McShane et al., 2024; Rovetta & Mansournia, 2024). A "small" P-value indicates that the observed result would be "rare" under A and $H_i$; however, this statement alone does *not* determine whether the model should be refuted in practice. The transition from model-based evaluation to real-world action requires integrating this information with external considerations, including subject-matter knowledge, causal plausibility, data quality, and the credibility of the underlying assumptions (Greenland, 2025; McShane et al., 2024; Rovetta, 2025; Neyman, 1977; Rovetta & Mansournia, 2024; Greenland, 2022; Box, 1980; Good, 1992; Hennig, 2010; Gelman & Loken, 2014; Rovetta, Mansournia, et al., 2025). Here one can see that probability is a model-defined numerical quantity (i.e., the long-run relative frequency under repeated realizations of the same experiment under A + $H_i$), whereas rarity is an interpretive judgment about how unusual that quantity is considered in context.

This leads to the notion of 'acceptability' as a decision-oriented concept. Acceptability reflects a judgment that combines data-model compatibility with external evidence and contextual considerations including usefulness and consequentiality (Good, 1992; Greenland & Hofman, 2019; McShane et al., 2024; Neyman, 1977; Rovetta & Mansournia, 2024). Thus, the evaluation of acceptability implicitly involves a loss structure (Greenland, 2021): what are the consequences of rejecting a model that is *at least* "adequate" given the established goal, and what are the consequences of retaining a model that is "inadequate" given the established goal?

In this regard, especially in medical settings where the stakes are extremely high, it is essential to emphasize that there is often *more than one* model that is *at least* reasonably compatible with the data - and thus potentially acceptable (Fisher, 1955; Greenland, 2025; Rovetta, Mansournia, et al., 2025). Within the same family of models defined by A, multiple $H_i$ may exhibit high degrees of compatibility with the observed data, and even alternative families of assumptions distinct from A may also be consistent with the same data. It is important to note that the aforementioned

"reasonableness" is not a property of models, but a human conclusion influenced by factors including competence, creativity, and moral values (Fisher, 1955; Gelman & Hennig, 2017; Good, 1992; Greenland, 2012, 2023a, 2025; Hennig, 2010; Rovetta, 2025; Rovetta, Biscetti, et al., 2026). For these reasons, from now on, instead of passive expressions such as 'given the assumptions', expressions that emphasize human agency will be used, such as 'according to the assumptions specified by the analyst'.

To evaluate any degree of acceptability, it can be useful to introduce the notion of refutational information. Suppose we observe a sequence of tosses and record the number of heads obtained. We define a model by combining a set of assumptions A - including independence and identical distribution of the tosses - with a composite target hypothesis $H_0$ stating that the coin is *not* biased in favor of heads (i.e., the probability of heads is less than or equal to 1/2). Within this model, the number of heads in 's' tosses follows a binomial distribution; in particular, the probability of observing 's' heads in 's' tosses is *at most* $1/2^s$ - where $1/2^s$ is the boundary case in which the probability of heads matches exactly that of tails (Cole et al., 2021). The P-value for such a one-tailed scenario is thus $p = \sup\{\Pr(S \geq s)\} = \sup\{\Pr(S = s)\} = 1/2^s$; this corresponds exactly to the probability of obtaining 's' heads in 's' tosses under A and the boundary case of $H_0$ in which the coin is not biased in either direction. In other words, the P-value is compared to the probability of obtaining 's' heads in 's' tosses under A and the specific case within the composite hypothesis $H_0$ in which the coin is not biased in either direction.

From the inverse transformation, one obtains a base-2 logarithmic relation that defines the S-value, $s = -\log_2 p$, which can be interpreted as the amount of information in 'bits' against the *acceptability* of the composed hypothesis $H_0$ (the coin is not biased toward heads), conditionally on the working model A defined by the analyst (McShane et al., 2024; Rafi & Greenland, 2020; Rovetta, 2025; Rovetta, Stovitz, et al., 2025; Rovetta & Mansournia, 2024). This coin-toss analogy can be naturally extended to evaluate the acceptability of any hypothesis (e.g., the hypothesis of a zero average effect of a medical treatment) under a given set of assumptions (Rafi & Greenland, 2020; Rovetta & Mansournia, 2024). The S-value allows one to express how much *refutational information* the data provide against the acceptability of a target hypothesis $H_i$ within a specified working model A, compared to the refutation information that 's' heads in 's' tosses of a coin provides against the acceptability of the hypothesis ‘the coin is not biased toward heads’ under the assumption ‘the coin toss is fair’.

## 5. S-VALUES: SCALE, INTERPRETATION, AND REFUTATIONAL INFORMATION

A key limitation of the P-value lies in the fact that it does *not* preserve information on the additive scale (Mansournia et al., 2022; Rafi & Greenland, 2020; Rovetta, 2025). For example, the change from 0.04 to 0.01 represents a substantially larger shift in evidential meaning than the change from 0.94 to 0.91, even though both differences are numerically equal. This asymmetry can be initially appreciated in qualitative terms by considering the multiplicative scale: in the first case, 0.04 is four times larger than 0.01, indicating a “substantial” change in the rarity of the result under the model; in contrast, 0.94 is only about 1.03 times larger than 0.91. However, since the multiplicative scale exaggerates differences, a more precise assessment is obtained by translating P-values into S-values - which address P-values dependence on location within the unit interval. Under such transformation, $p = 0.01$ corresponds to approximately $s = -\log_2 0.01 = 6.64$ bits, while $p = 0.04$ corresponds to about 4.64 bits, yielding a difference of 2 bits;[1] by contrast, $p = 0.91$ corresponds to about 0.14 bits and $p = 0.94$ to about 0.09 bits, with a difference of only about 0.05 bits.

Because it is defined on a logarithmic base-2 scale, equal increments in the S-value correspond to equal increases in refutational information (provided by the data against the acceptability of the targeted hypothesis, given the assumptions fixed by the analyst). For instance, the difference between $s = 9$ bits and $s = 10$ bits represents the same additional amount of information (1 bit) as the difference between $s = 19$ bits and $S = 20$ bits. An additional advantage of the S-value is that it allows probabilities to be mapped onto a concrete and familiar physical experiment, namely repeated coin tossing. This provides an intuitive benchmark grounded in everyday experience and helps contextualize the degree of rarity of the observed data in relation to the specific research question and decision context, supporting more informed judgments about the acceptability of the underlying model.

[1] A faster way to compute differences is to exploit the properties of logarithms, so that $\Delta s = \log_2(0.04/0.01) = 2$ bits.

In some work - including my own (Rovetta, Stovitz, et al., 2025) - this refutational information was often described as being directed against the model itself, rather than against its acceptability. Upon closer inspection, such formulation is imprecise and risks conveying a misleading message, namely that "atypical" results under A + $H_i$ (i.e., small P-values and thus large S-values) can provide evidence against a model that is, by construction, assumed to be (provisionally) true. Neither the P-value nor the S-value can be interpreted as providing evidence about the truth or falsity of the model in an epistemic sense (Greenland, 2008): the evaluation of acceptability lies *outside* the "fictional small world" induced by the model, i.e., it lies in the real-world context in which decisions are made. Accordingly, the S-value should *not* be interpreted as evidence against the model per se: rather, it contributes to informing judgments about how much the model remains acceptable, without being sufficient, on its own, to determine that human decision.

A more precise interpretation is that S-values characterize the degree to which the observed data depart from what would be expected under the model assumed by the analyst. The evaluation of whether this departure renders the model "unacceptable" depends on contextual considerations based on the specific scope of the analyst - shaped by their background knowledge and their intention (Rovetta, Stovitz, et al., 2025). In other words, the S-value does *not* quantify evidence against the epistemic status of the model assumed for the calculation; rather, it quantifies a specific form of refutational information that becomes part of the broader body of information the analyst uses to judge whether the model remains acceptable for the intended practical purpose.

For example, observing 2 heads in 2 tosses is typically regarded as weak refutational information against the acceptability of the hypothesis 'the coin is not biased toward heads'. This is because, in ordinary real contexts, such an outcome is not perceived as "rare" even when the plausibility of the model defined by A and $H_0$ is high - in light of our experience and cognitive perceptions. So, when the experiment is conducted under routine conditions - where assumptions such as toss independence are considered reasonable by the analyst - 2 heads in 2 tosses are generally insufficient to motivate refuting the model as an adequate description of the real process. By contrast, observing 10 heads in 10 tosses may be considered "sufficiently" rare under A together with $H_0$ to prompt a different judgment. In that case, alternative explanations - such as the possibility that the coin is in fact biased - may appear more consistent with the observed data. As a result, the hypothesis $H_0$ may be deemed unacceptable, in the sense that it is no longer considered a convenient representation - or even idealization - of the *real* phenomenon under investigation given the analyst's scope.

Nonetheless, it is important to stress that overall support for a hypothesis requires a restriction of plausible alternatives (Greenland, 2023b; Rovetta, Mansournia, et al., 2025); for example, in this case we must exclude the possibility that the person tossing the coin is a magician capable of influencing the outcome at will (e.g., by using his craft of distracting the audience's attention), even if the coin itself is not biased at all.

## 6. LOSS ACCEPTABILITY

There is no universal or context-free scale for evaluating the degree of refutational information. Any such evaluation is necessarily conditional on the *loss function*, that is, on how the consequences of decisions are weighted by the analyst across possible states of the world (Good, 1992; Greenland, 2021, 2022; Neyman, 1977; Rovetta & Mansournia, 2024). The loss function encodes the costs associated with different types of errors *as evaluated by the specific analyst*, conditionally on their knowledge, moral values, perceptions, and goals. These costs are not necessarily symmetric and strongly depend on the specific practical context. To illustrate a case of marked asymmetry, consider a metal detector: the cost of a false negative (e.g., a person is carrying a gun but the alarm does not go off) is much higher than the cost of a false positive (e.g., the person is not carrying a gun but the alarm goes off). In this case, even "weak" evidence against the acceptability of the null hypothesis 'the person is not carrying a gun' may be more than sufficient to trigger the action 'search the person.' So, the same amount of refutational information may lead to different decisions depending on how severe the consequences of being wrong are judged to be.

As another stylized example, suppose you are offered the following bet: if you wager that a coin is biased toward heads and this hypothesis turns out to be correct, the gain is 100 dollars; if the hypothesis turns out to be incorrect (i.e., if the coin is not biased toward heads), you lose 0 dollars. The question is: what is the number ‘s’ of heads in ‘s’ tosses that you should observe in order to take the bet? If the ultimate goal is to earn money, the optimal decision is then trivial: you should accept the bet regardless of the observed data, since there is no downside cost. Now suppose that losing the bet entails a cost of 5 dollars. In this setting, the personal decision becomes sensitive to the degree of refutational information provided by the data. One might judge that observing 3 heads in 3 tosses - corresponding to a “modest” level of rarity under the assumption of no bias toward heads - is sufficient to justify the risk. However, if the potential loss increases to 100 dollars, matching the possible gain, a much stronger degree of refutational information would typically be required. For instance, one might require observing at least 5 heads in 5 tosses before considering the bet acceptable.

This dependence becomes even more pronounced in sensitive settings such as medicine (Fraiman et al., 2025; Greenland, 2021; Rovetta, Stovitz, et al., 2026; Rovetta & Mansournia, 2024). For example, consider a setting in which the causal structure underlying a treatment is clear (e.g., the biological mechanisms are well understood) and the study design is optimal. When assessing a hypothesis concerning a relatively mild outcome, such as a non-negligible increase in the frequency of headaches, a moderate level of refutational information (e.g., 4 bits) might be considered sufficient. By contrast, when evaluating severe outcomes, such as anaphylaxis, a substantially higher degree of refutational information would typically be required before deciding to refute that costly hypothesis. In this regard, it is essential to recognize that the decision problem becomes even more complex, as it does not reduce to a binary accept-or-reject choice, but instead requires selecting among several therapeutic alternatives - whose benefits and side effects may differ across several practical aspects, including dosage, timing of administration, and treatment intervals.

Loss-based reasoning should also guide adjustments for multiple comparisons (Fraiman et al., 2025; Greenland, 2021; Greenland & Hofman, 2019; Rovetta, Stovitz, et al., 2026). Specifically, multiplicity adjustment should not be applied automatically merely because several tests are performed; whether adjustment is warranted, and how stringent it should be, depends on the relevant loss structure and inferential or decision goal - both of which should be openly and fully described before conducting the data analysis, based on the analyst's prior knowledge and objectives (Greenland & Hofman, 2019; Neyman, 1977; Rovetta & Mansournia, 2024). Moreover, the family of tests to be adjusted should be defined by those that jointly contribute to the same conclusion or decision, rather than indiscriminately including every test performed in an analysis.

## 7. ON SURPRISAL AND SURPRISE

S-values are often referred to as *surprisals*, a term that is frequently used alongside the more general notion of *surprise* - especially when illustrated through analogies such as observing ‘s’ heads in ‘s’ tosses under the assumption that the coin is not biased toward heads (Cole et al., 2021; Greenland et al., 2022; McShane et al., 2024; Rafi & Greenland, 2020; Rovetta, 2025). While this analogy is useful, it can also generate ambiguity unless the notion of “surprise” is carefully specified. The term “surprise” can suggest a deviation from what is expected, but its meaning varies depending on the framework in which it is used. In everyday language, surprise is typically associated with a *psychological reaction* to an event perceived as unusual (Cambridge University Press, n.d.). In statistical contexts, however, at least two distinct notions should be clearly distinguished: a Shannon-type (information-theoretic) notion and a Bayesian notion.

In the Shannon sense, surprisal is defined purely in terms of probability within a specified model (Cole et al., 2021). An event is “surprising” to the extent that it has “low” probability under the assumed system.[2] The S-value quantifies how unexpected a certain outcome (e.g., obtaining a test statistic as or more extreme than that observed in the experiment) is within the “small fictional world” induced by the chosen model - which is *provisionally assumed true*. This does *not* imply any claim about the truth of that world or require any revision of beliefs. By contrast, in a Bayesian framework, surprise is tied to belief updating (Baldi, 2002). An observation is surprising to the extent that it leads to a revision of prior beliefs, as formalized through the posterior distribution. In this sense, surprise is inherently epistemic: it reflects a change in what is considered plausible after observing the data - even if, in practical terms, usefulness is both a more achievable and interesting goal; see (Box, 1976).

[2] Adjectival terms such as “surprising” and “low” are placed in quotation marks precisely because these evaluations depend on loss acceptability; they are not universal.

A simple example helps clarify this distinction. Suppose a football player widely regarded as poorly skilled performs an extraordinary bicycle kick and scores a remarkable goal. This event may be perceived as “highly surprising” in the sense that it is “rare” given one’s expectation, that is, it *diverges* “a lot” from one’s expectation. However, this does *not* necessarily imply a revision of one's overall judgment about the player's ability: the event may instead be interpreted as an isolated occurrence or a matter of chance, without altering the prior belief that the player is poorly skilled. Analogously, within the frequentist framework, observing a result with a high S-value indicates that a certain outcome is rare under the assumed model.

In this regard, it should be noted that Frequentist and Bayesian approaches are therefore *not* in opposition, but can be better understood as complementary components of a broader inferential framework (Box, 1980; Greenland, 2008; Greenland & Hofman, 2019). Indeed, decisions about loss acceptability inevitably incorporate prior considerations and beliefs - including potential revisions (Greenland, 2022; Rovetta & Mansournia, 2024); and, as emphasized by Box (Box, 1980), model assumptions themselves can be viewed as fixed priors.

A distinct interpretation of surprise can be derived from its ordinary usage: an event is surprising if it *produces* (a sense of) surprise (Cambridge University Press, n.d.). One may speak of a “surprising” hypothesis not because the hypothesis itself is “rare”, but because, when evaluated against the observed data, it generates a certain degree of surprise (Rovetta, Msaouel, et al., 2025). Operationally, one can fix the observed data and the set of assumptions A, and then vary the hypothesis under consideration ($H_1$, $H_2$, …) to assess how much refutational information each hypothesis produces given A and the data (Msaouel, 2024; Rovetta, Msaouel, et al., 2025; Rovetta, Stovitz, et al., 2025). Under this interpretation, the surprisal ‘s’ becomes a function $s = s(H_i; A, D)$ of the hypothesis $H_i$ given the observed data D and the assumptions in A. Crucially, it does *not* refer to the rarity of the hypothesis itself (which is assumed 100% true within the model), but to the amount of refutational information (in bits) that the hypothesis yields *against* *its real-world acceptability* when confronted with the observed data, holding the assumptions A constant (Rovetta, 2025; Rovetta, Msaouel, et al., 2025).

Within this perspective, the surprise does not arise inside any single model, but emerges from a comparison process. It is experienced at the level of the analyst, who evaluates how each hypothesis fares when confronted with the data. The numerical outputs produced within each model (e.g., the S-value) serve as mere indicators, quantifying the degree of refutational information associated with each hypothesis under the assumptions specified by the analyst; but the practical interpretation of that information remains anchored in the real world, through the *overall* acceptability evaluation.

To illustrate, suppose we observe a marked clinical improvement in a patient following administration of a treatment (the observed data D). The set of assumptions A encodes the idealized data-generating process, including correct outcome measurement, patient adherence, absence of major confounding factors, and a stable clinical condition unrelated to external interventions. Now consider two alternative hypotheses: under $H_0$, the treatment has zero effect; under $H_1$, the treatment has a specific "large" effect. Holding D and A fixed, $H_0$ becomes "very" surprising when evaluated against the data, whereas $H_1$ does not (as it is "very" compatible with D). Importantly, the sense of surprise associated with $H_0$ does not arise because $H_0$ is intrinsically "rare" or "unlikely"; rather, it emerges from the mismatch between the expectations implied by A together with $H_0$ and the observed data. In other words, *given the assumptions* in A and compared to $H_1$, $H_0$ generates a higher amount of refutational information against its own real-world acceptability when confronted with the data, despite being treated as provisionally true within the model $H_0$ + A.

It is worth noting that previous works - including my own (Rovetta, Msaouel, et al., 2025) - have made use of the notion of such a Shannon-type surprise. However, upon closer consideration, this interpretation could have criticalities. The main concern arises from the well-known *inversion fallacy*, whereby the probability of an outcome under a hypothesis is mistakenly interpreted as the probability of the hypothesis given the outcome (Gigerenzer, 2018; Greenland et al., 2016; Rovetta & Mansournia, 2025; Wasserstein & Lazar, 2016). The former is a frequentist quantity, defined within the assumed model; the latter is a fully Bayesian construct, which requires the explicit specification of additional prior distributions beyond the standard modeling assumptions (Greenland et al., 2016). In this light, the above concept of Shannonian surprise is perfectly defensible, but its use carries a non-negligible risk of encouraging such misinterpretations - especially in applied fields such as clinical research, where statistical concepts are often communicated to a broader audience.

For this reason, although it is important to clarify the proper interpretation of this notion wherever it appears, I do *not* recommend its routine use in contexts that are particularly prone to highly consequential misunderstanding. Its application may be more appropriate in technically oriented statistical work, where the audience is equipped to distinguish clearly between model-based quantities, decision-making, and epistemic probabilities.

## 8. COMPATIBILITY, REPORTING, AND THE LIMITS OF SINGLE-STUDY INFERENCE

Under the present framework, it is legitimate to refer to the compatibility of hypotheses with the data - especially when comparing several target hypotheses while holding the data and the other assumptions fixed in order to inform a real-world acceptability evaluation. In such cases, the analyst may examine how the same observed data align with different hypotheses (e.g., $H_0$, $H_1$, $H_2$, …), thereby assessing their relative compatibility within a common modeling structure - according to the rules imposed by that structure (e.g., the test statistic under the chosen working model) (Amrhein, Greenland, et al., 2019; Greenland, 2023b; Mansournia et al., 2022; Rovetta, Piretta, et al., 2025). However, this formulation also carries a potential risk of misinterpretation: referring to the ‘compatibility of a hypothesis with the data’ may inadvertently suggest an evaluation of the hypothesis itself, thereby encouraging the inversion fallacy. For this reason, in settings where such confusion is very likely and consequential, it may be preferable to adopt an even more cautious formulation and speak instead of the compatibility of the data with a given hypothesis under A.

Such an oriented ‘data-model’ compatibility phrasing may better reflect the frequentist logic of the approach: the analyst considers the observed data and evaluates their consistency under each hypothetical model defined by A together with $H_i$, without attributing probabilistic statements to the hypotheses themselves. This perspective preserves conceptual clarity while reducing the risk of inadvertently importing interpretations that belong to a different inferential framework.

As stressed by previous research (Amrhein, Trafimow, et al., 2019; Bann et al., 2024; Berner & Amrhein, 2022; Greenland et al., 2022; Hawkins & Samuels, 2021; Higgs & Amrhein, 2025; Mansournia & Nazemipour, 2024; McShane et al., 2019; Rafi & Greenland, 2020; J. Rothman, 1978; K. J. Rothman, 1986; Rovetta, Mansournia, et al., 2025; Ting & Greenland, 2024; Wasserstein et al., 2019; Wasserstein & Lazar, 2016), only a minority of empirical investigations should be conceived with the explicit goal of formal, conclusive decision-making (even if based on fully specified loss structures). In most research settings - particularly in clinical and epidemiological studies - the primary objective is (or should be) to describe and summarize the relationship between data and hypotheses in a transparent and interpretable manner (Greenland, 2023b, 2023a). This caution is essential to avoid overinterpretations that lead to premature decisions (based on evidence that is not yet sufficiently stable or robust), resulting in waste of resources, reduced public trust, and, in sensitive fields such as medicine, harm to the most vulnerable stakeholders (Amrhein, Trafimow, et al., 2019; Bann et al., 2024; Rafi & Greenland, 2020; Ting & Greenland, 2024).

In such cases, a less conditional approach is more appropriate (Rafi & Greenland, 2020; Rovetta & Mansournia, 2024). The analysis proceeds by evaluating the observed data under specified assumptions A and hypotheses H, and by reporting measures of compatibility or refutational information that characterize this relationship. The aim is not to deliver definitive judgments about acceptability or to prescribe decisions, but to provide coherent and calibrated summaries that can be meaningfully interpreted and compared across studies. Such summaries are especially valuable in the broader process of evidence synthesis. Systematic reviews and meta-analyses rely on the availability of interpretable and comparable statistical outputs across individual studies; accordingly, emphasizing transparent descriptions of data-model relationships is more aligned with the cumulative nature of scientific inference than attempting to embed complex and often uncertain loss structures within single studies (Greenland, 2025; Higgs & Amrhein, 2025; Rovetta, Stovitz, et al., 2025).

## 9. THE CONTINUITY BETWEEN FREQUENTIST AND BAYESIAN REASONING

The contrast between frequentist and Bayesian approaches often rests on a wrong opposition: that Bayesian analyses rely on priors or (subjective) beliefs, whereas frequentist analyses are free from such epistemic commitments (Gelman & Hennig, 2017; Greenland, 2008). This description is misleading in several respects. As discussed in paragraphs 1-8, a frequentist analysis does not begin from nowhere: before any P-value or S-value can be computed, the analyst must select at least one working model $A_i$ from among the many available or conceivable alternatives (e.g., selecting a regression model among many alternatives differing in covariate selection, functional form, and the inclusion of additive effects or multiplicative interaction terms). That choice is neither random nor epistemically empty: it reflects a prior judgment that $A_i$ is sufficiently credible, useful, and scientifically defensible for the purpose at hand (Box, 1980; Cox & Hinkley, 1979; Fisher, 1955; Gelman & Loken, 2014; Greenland, 2022, 2023b; Rovetta & Mansournia, 2024). The analyst may justify this judgment qualitatively, for example by arguing that $A_i$ "adequately" reflects the data-generating structure; but whether expressed qualitatively or quantitatively, such a judgment still functions as a prior commitment.

One way to make this feature more explicit is to imagine an idealized model-selection process over a "set" (or, better, a space), of *candidate families of assumptions* $\{A_i\}$ - see (Cox & Hinkley, 1979). Within that perspective, the prior component can be given an idealized repeated-sampling reading: one may imagine a large sequence of "epistemically equivalent pre-analytic settings" in which, under similar background knowledge and practical goals, candidate assumption families are evaluated for their acceptability as working models (Rubin, 2020). The prior assigned to the specific $A_i$ under examination is represented by the limiting relative frequency with which $A_i$ would be judged acceptable across those hypothetical repetitions. This reading is also coherent with a constructivist perspective (Hennig, 2010): $Pr_C(A_i)$ may be viewed *not* as a probability of literal "truth", but as an estimated measure of *intersubjective agreement* within a certain class C of "reasonable" analysts - see (Greenland, 2025). This can be understood in two related senses: i) as convergence among different analysts assigning prior credibility to $A_i$ under similar background conditions, and ii) as the limiting relative frequency with which future analysts would assign comparable prior credibility, or judge $A_i$ acceptable, across hypothetical repetitions of similar pre-analytic settings.

Within that perspective, choosing a specific $A_i$ may be understood as requiring that its prior credibility exceed some minimum required level of satisfaction δ; in other words, the analyst requires that $Pr_C(A_i) \geq \delta$, where δ represents a context-dependent condition of acceptability. In many applied analyses, this "indicative, coarse threshold" is left implicit, and the analyst merely states that the assumptions are considered "reasonable" given their current state of knowledge and perception (Greenland, 2023b, 2025). Yet these verbal formulations do not remove the prior component; they only leave it unformalized (Box, 1980). The selection of $A_i$ therefore already encodes a judgment about which embedding models are sufficiently credible to deserve conditional analysis.

Bayesian approaches are not entirely separate inferential systems: they can be viewed as extensions of conventional frequentist approaches that retain similar data-generating assumptions while adding further assumptions - in the form of specific prior distributions over parameters (Box, 1980; Greenland, 2008). For example, an analyst may judge, based on their own interpretation of available information, that very large treatment effects are implausible; thus, they may assign most prior probability to a range they regard as sufficiently plausible and relatively little probability to values they consider extreme. Such prior probabilities should thus be understood as explicit representations of the analyst's pre-data judgments, which may be informed by external evidence and calibrated against repeated experience. Under a calibration interpretation, an analyst who repeatedly assigns probability 0.80 to comparable propositions would ideally expect roughly 80% of them to be borne out over a suitably defined long run. Even within a pure frequentist framework, repeated observations of "low" agreement between the data and a model - especially when combined with biological, causal, clinical, and other external evidence - can lead the analyst to revise the prior credibility (implicitly) assigned to that model and eventually reject it as an acceptable representation of the process under study.

Major differences between frequentist and Bayesian approaches therefore lie in how the resulting information is interpreted and used. Typical Bayesian analysis combines the likelihood function with a prior distribution to obtain a posterior probability distribution over the parameter. Thus, under the assumptions defining such a Bayesian procedure, a 95% posterior interval assigns 95% posterior probability to the parameter lying within that interval, conditional on both the prior distribution and the assumed data model (Box, 1980; Greenland, 2008). However, in parallel with the frequentist framework, this quantity does not provide information on the credibility, adequacy, or reliability of the procedure that generated it. Such an interval may therefore be more cautiously described as a Bayesian compatibility interval (McElreath, 2020): a range of parameter values that remain reasonably compatible with the data and prior information according to the specific Bayesian procedure specified by the analyst - without implying any unwarranted "credibility".

As a final caution, Bayesian analyses may be especially vulnerable to the pre-direction of results because sufficiently informative priors can strongly constrain - and in extreme cases largely predetermine - posterior conclusions (Spiegelhalter et al., 2003). This risk is particularly relevant when priors are based on expert opinion, since such judgments may inherit distortions from the literature, including publication bias, selective reporting, statistical misinterpretations, entrenched positions, and conflicts of interest. Spiked priors assigning substantial probability to the null hypothesis are a notable example, as they can generate strong posterior support for the null mainly because that support has already been built into the prior (Casella & Berger, 1987).

CONFLICT OF INTEREST STATEMENT

The author declares no conflict of interest.

DATA AVAILABILITY STATEMENT

The data supporting the findings of this study are derived from publicly available PubMed records. The search keywords and corresponding summary counts are provided in the Supplementary Material, available at: https://doi.org/10.6084/m9.figshare.33602353.

ACKNOWLEDGEMENTS

The author thanks Marcel Miché for carefully reading an earlier version of the manuscript and for providing helpful comments and suggestions.